# Analysis of Fracture and Fatigue using a Simplified Stress Wave Unloading Model and Lagrangian Mechanics – Version 2

Clive Neal-Sturgess
Emeritus Professor of Mechanical Engineering
University of Birmingham.

## Introduction

Many people are aware of the theory of elastic fracture originated by AA Griffith in 1920 (Griffith 1921), and although Griffith used the "*theorem of minimum potential energy*" (his italics) most people seem unaware of the broader implications of this theorem. If it is set within its classical mechanics roots, it is clear that it is a restricted form of a Lagrangian, and conforms to Euler-Lagrange as shown below.

In "advanced" texts on fracture Freund (Freund 1990), Kanninen and Popelar (Kanninen and Popelar 1985), and Williams (Williams 1984) cracks are treated as dynamic entities, and the role of stress waves is clearly articulated, also see Kolsky (Kolsky 1963). However, in most non-advanced texts on fracture and fatigue the role of stress waves are either not included or not emphasised, often leading to a possible misunderstanding of the fundamentals of fracture. The position adopted here is that all cracks are dynamic entities, driven by a stress wave unloading model. Fatigue cracks are simply slow-moving cracks, as they only travel a short distance each time they propagate, and so do not need inertial corrections.

What is done here is to extend Griffith's approach by setting it within the concept of Stationary Action (Hamilton's Principle) (Penrose 2007), and introducing a simplified stress wave unloading model, which connects the energy release mechanism with the stress field. This leads to a definition of a dynamic stress intensity factor for long running cracks, and this model is then applied to fatigue of perfectly elastic and elastic-plastic materials to include crack tip plasticity.

The results for the Griffith crack is retrodiction, to establish the validity of the methods used. The extension to the dynamic case and fatigue gives significant new results. For fatigue for elastic-plastic materials the influence of the maximum stress in the cycle as a fraction of the yield stress, called the yield stress ratio (YSR) is identified as a significant factor 229-260. The new form of the fatigue crack growth relationship derived answers many of the long standing questions about the Paris Law (Hertzberg 1976).

### Part I The Griffith Crack

Griffith took the case where a crack is inserted into an infinite stressed body and results in a circular unloading zone as shown in Fig 1.

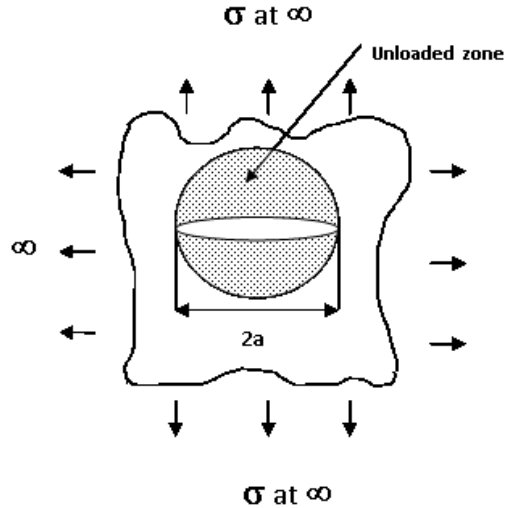

Fig. 1.1 The Griffith Crack (after Benham & Crawford)

Griffith derived an instability condition for crack propagation using the theorem of minimum energy, he called it 'G$_c$', this was confirmed by Sneddon (Sneddon 1946) who conducted a stress analysis of an embedded crack, and he also determined the same instability condition.

In1957 Irwin (Irwin 1957) also determined the Griffith instability condition and defined it as "the critical strain energy release rate" (K$_c$), and in 1958 Irwin (Irwin 1958) made the connection between the 'K' dominated stress field, and 'G$_c$' as:

$$G_c = \frac{K_c^2}{E} \tag{1.1}$$

This is the well-known Griffith-Irwin fracture criteria. To recast this in terms of stationary Action, consider the following.

If a crack is embedded, and the boundaries of the plate are at infinity, then the system is "closed ", that is $\frac{dU_T}{dt} = 0$, and so Lagrangian mechanics apply. Consider the quantity:

$$S \equiv \int_{t_1}^{t_2} L(x, \dot{x}, t) \tag{1.2}$$

where $S$ is called the *Action*, and $L$ is a Lagrangian.
Consider a function $x(t)$, for $t_1 < t < t_2$, which has fixed endpoints (that is, $x(t_1) = x_1$ and $x(t_2) = x_2$), but is otherwise arbitrary. A function $x(t)$ which yields a stationary value of $S$ (a stationary value is a local minimum, maximum, or saddle point) satisfies the Euler-Lagrange equation:

$$\frac{d}{dt}\left(\frac{\partial L}{\partial \dot{x}}\right) - \frac{\partial L}{\partial x} = 0 \tag{1.3}$$

where usually $L = T - V$ (1.4)
and $\quad$ V = the gravitational potential energy
$\quad$ T = the kinetic potential energy

However, for the fracture problem where the crack is propagating horizontally, it can be assumed to have a constant value of the gravitational potential, and the energy transfer is the strain energy in the body being converted into new crack surfaces. For this problem, a Lagrangian can be defined as:

$$L = H - M \quad (1.5)$$

where: $\quad$ H = the energy required to create new crack surfaces
$\quad$ M = the strain energy released from the body

This is a similar technique to the derivation of Sneddon (1946), who again used the principle of minimum potential energy.
Assuming that a crack will propagate when the Euler-Lagrange equation is satisfied (*i.e.* that the Action is stationary), let $x = a$, where $a$ is the semi crack length, see Fig. 1 (a Griffith crack), then the energy required to create the crack surfaces per unit thickness is:

$$H = 2\gamma a \quad (1.6)$$

where $\gamma$ is the energy per unit crack length.

In the static Griffith model outlined above, and in current interpretations (Benham and Crawford 1996), the unloaded zone is considered to be a circumscribed circular area equal in diameter to the length of the embedded crack, i.e. 2a, which is a considerable idealisation and although a good approximation at the time of Griffith, does not bear any relationship to the actual physics of the situation.

Considering the physics of fracture, the crack tip is under a tensile stress $\sigma_*$, and when the tip of an atomically sharp crack separates, in an idealised model, the atomic bond at the crack tip separates and stress at the crack tip falls to zero. This "stress drop", as it is known in Tectonophysics, can be approximately equal to the stress at the crack tip (Kanamori 1972). This phenomena was filmed in 1959 by Schardin (Schardin 1959) using high speed Schlieren photography, as shown in Fig. 2: the stress waves can be seen emanating from the crack tip. It is these unloading waves which release the strain energy from the body, and make it available for generating new crack surfaces.

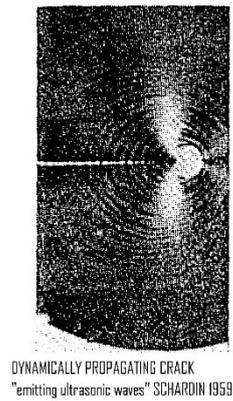

Fig. 1.2 A dynamically propagating crack, after Schardin (Schardin 1959)

If the material is perfectly elastic, and the stress field at the tip of the crack is K dominated, then after the local stress state at the crack tip can be described as in Fig. 3.

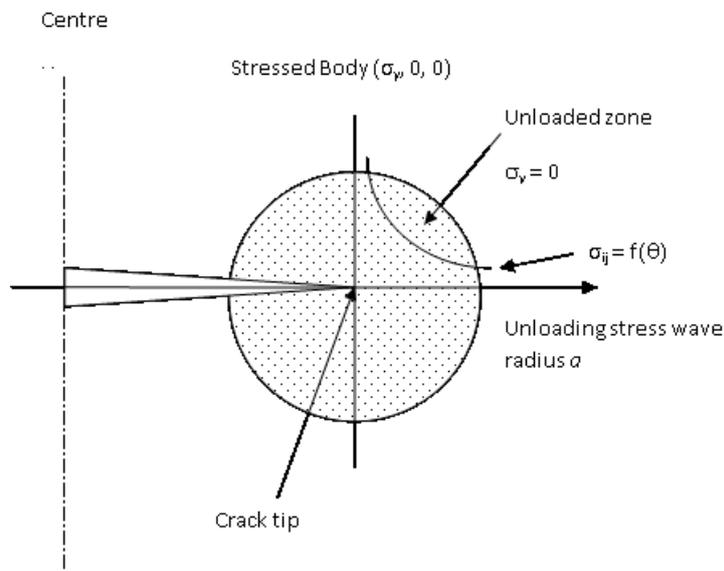

Fig. 1.3 Schematic of an unloading stress wave at the tip of a crack.

Then

$$\sigma_{ij} = \frac{K}{\sqrt{2\pi r}} f(\theta) = \sigma_* \qquad (1.7)$$

Letting $f(\theta) = 1$, then $r = \dfrac{K^2}{2\pi \sigma_*^2}$

The energy released from the body by the unloading stress wave may then be modelled as, for a unit thickness:

$$M' = \frac{\sigma_*^2}{2E}(vol) = \frac{\sigma_*^2}{2E}(\pi r^2) \tag{1.8}$$

Where $r$ is the radius of the unloading stress wave. However, the stress $\sigma_*$ is a function of $r$ and so the average stress is $\bar{\sigma}_* = \frac{K}{\sqrt{2\pi\bar{r}}}$. For the Griffith case it is necessary to take $\bar{r} = a/2$ and the Lagrangian then becomes:

$$L = 2\gamma a - \frac{K^2 a}{E} \tag{1.9}$$

So $\quad \frac{\partial L}{\partial a} = 2\gamma - \frac{K^2}{E} = 0 \tag{1.10}$

Hence $\quad 2\gamma E = K^2 \tag{1.11}$

which is the Griffith-Irwin equation derived without recourse to two separate equations defining the Griffith and Irwin criteria separately, and is a major simplification of existing derivations. It arises because modelling the physics of the situation by a stress wave unloading mechanism connects the energy released from the body with the K-dominated stress field ahead of the crack.

**Part II A Dynamic Stress Intensity Factor for a moving crack in a perfectly elastic material**

For a moving crack an inertial correction is necessary, as when the crack is moving energy is required to displace the crack flanks normal to the direction of crack propagation. Mott (Mott 1948) was the first to derive an inertial correction in terms of the main parameters considered here. The integrated form of Mott's inertial correction (Ewalds and Wanhill 1983) was taken as:

$$T^+ = \frac{k}{2}\rho a_i^2 V^2 \left(\frac{\sigma}{E}\right)^2 \tag{2.1}$$

where:  k = an undetermined constant
 $\rho$ = density
 $a_i$ = instantaneous crack length
 V = crack velocity

This analysis was criticised (Kanninen and Popelar 1985) because Mott considered that all the excess energy released from the body went into the kinetic energy of the crack, and the crack acceleration ($\ddot{a}$) was zero. A subsequent analysis by Roberts and Wells (Roberts and Wells 1954) showed that:

$$V = 0.38 C_o \left(1 - \frac{a_o}{a}\right)^{1/2} \tag{2.2}$$

where $C_0$ is the velocity of sound in the material.

Now Mott's assumption of $\ddot{a} = 0$ is reasonable if the *terminal* velocity of the crack is being considered, as cracks initially accelerate very quickly and finally approach their terminal velocity slowly (Kanninen and Popelar 1985). Furthermore, carrying Mott's analysis to its final result gives:

$$\hat{V} = \dot{a} = \sqrt{\frac{\pi}{k}} C_o \tag{2.3}$$

Comparison with equation 2.2 gives:

$$\sqrt{\frac{\pi}{k}} = 0.38$$

Hence $k \cong 6\pi$ (2.4)

Mott's correction can also be expressed as:

$$T^+ = \frac{D\sigma^2 \pi a^2}{E}\left(\frac{\dot{a}}{V_L}\right)^2 \tag{2.5}$$

where $D = \frac{k\rho V_L^2}{2\pi E}$ and for a long crack $a_i \approx a$

Now, in the presence of crack inertia, less of the energy released will go into forming new crack surfaces, so the Lagrangian for this case gives:

$$L = \left[2\gamma a - \frac{D\sigma^2 \pi a^2}{2E}\left(\frac{\dot{a}}{V_L}\right)^2\right] - \frac{\sigma^2 \pi a^2}{2E} \tag{2.6}$$

where the terms for the energy flowing into the crack are bracketed to emphasise the sign of the inertial correction. Therefore, applying Euler-Lagrange to obtain the equation of motion of the crack:

$$\frac{\partial L}{\partial \dot{a}} = -\frac{2D\sigma^2 \pi a^2 \dot{a}}{EV_L^2} \tag{2.7}$$

$$\frac{\partial}{\partial t}\left(\frac{\partial L}{\partial \dot{a}}\right) = -\frac{2D\sigma^2 \pi}{EV_L^2}\left(a^2 \ddot{a} + 2a\dot{a}^2\right) = -\frac{4D\sigma^2 \pi a}{E}\left(\frac{\dot{a}}{V_L}\right)^2 \text{ by letting } \ddot{a} = 0 \tag{2.8}$$

$$\frac{\partial L}{\partial a} = 2\gamma - \frac{2D\sigma^2 \pi a}{E}\left(\frac{\dot{a}}{V_L}\right)^2 - \frac{\sigma^2 \pi a}{E} \tag{2.9}$$

Applying Euler-Lagrange gives:

$$2\gamma E = \pi \sigma^2 a [1 - 2D\left(\frac{\dot{a}}{V_L}\right)^2] \qquad (2.10)$$

Assuming that the energy required for new crack surfaces is independent of the crack velocity then $2\gamma E = K_{1C}^2$ and letting $\pi \sigma^2 a = K_{1d}^2$, then

$$K_{1d} = \frac{K_{1c}}{[1 - 2D\left(\frac{\dot{a}}{V_L}\right)^2]^{\frac{1}{2}}} \qquad (2.11)$$

This is a good approximation to the dynamic stress intensity factor of a moving crack (see Figs 2.1 and 2.2) and equation (2.11) is of the general form found by Willis (Willis 1967), Freund (Freund 1990) and Kanninen and Popelar (Kanninen and Popelar 1985), but not in terms of the stress intensity factor, as done here. Again, these results show the validity of the present analysis.

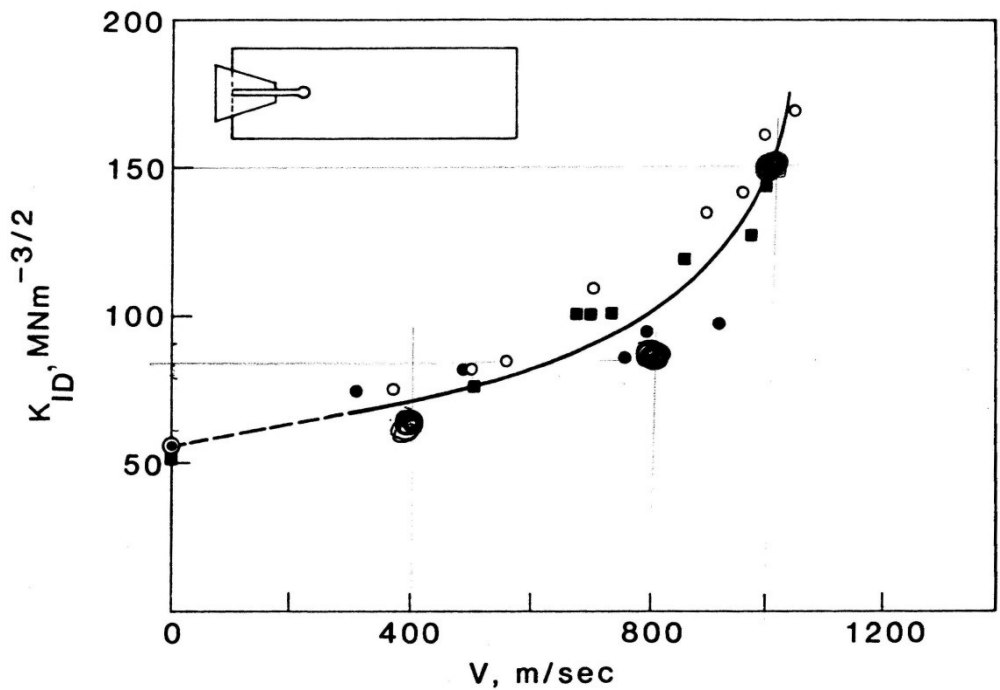

Figure 4.16  Dynamic fracture toughness values for 4340 steel.

**Fig. 2.1 Dynamic Stress Intensity Factor (from Kanninen & Popelar).**

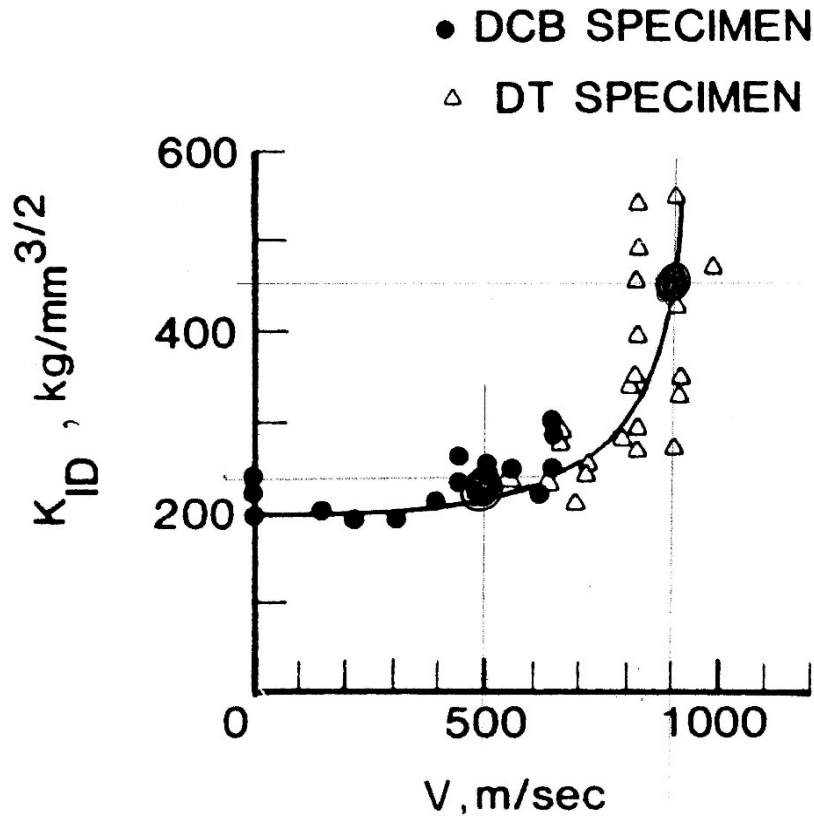

**Fig. 2.2 Dynamic stress intensity factor (from Kanninen & Popelar).**

**Part III Fatigue**

Fatigue was not addressed by Griffith, and this is the extension of his approach, embodied in a Lagrangian, which is the main purpose of this work. For a perfectly elastic material the system is conservative, and so a standard Lagrangian can be applied. As most materials used in situations where fatigue may arise are ductile then it is necessary to take crack tip plasticity into account. Extended Lagrangians can be used for non-conservative systems provided that the surface tractions are taken into account (Goldstein 1969). However, as here the domain boundaries are set at infinity, there will not be any effects of surface tractions and so a standard Lagrangian is used. This is a so-called "equilibrium solution" by Freund (Freund 1990), as overall the body is in static equilibrium.

**Perfectly Elastic Material Plane Stress**

For a fatigue crack, which is idealised as propagating a small distance in each cycle (see Fig. 3.1) along a curved crack front, as is often found (Roebuck 1992), the energy

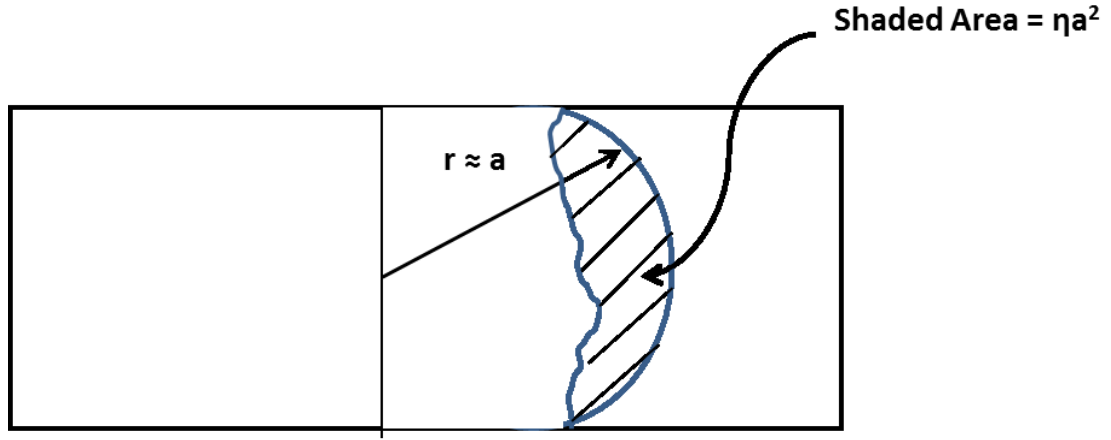

**Fatigue crack growth model**

FFig 3.1 Fatigue crack growth model

required to form the new crack surfaces can be expressed as:

$$T = 2\gamma\eta a^2 \tag{3.1}$$

where $\eta \ll 1$, then

$$L = 2\eta\gamma a^2 - \frac{K^2 a}{E} \tag{3.2}$$

$$\frac{\partial L}{\partial a} = 4\eta\gamma a - \frac{K^2}{E} = 0 \tag{3.3}$$

Assuming that crack extension takes place each cycle, which is a major idealisation (Scruby 1985), then let $\eta a = \frac{da}{dn}$, and taking for fatigue $K = \Delta K$, as is done in most cases, gives:

$$\frac{da}{dn} = \frac{\Delta K^2}{2 K_c^2} \tag{3.4}$$

where $K_c^2$ is the critical strain intensity factor for plane stress.

Such results have been derived before (Lardner 1967; Pelloux 1970; Pook and Frost 1973; Irving and McCartney 1977), but by using other models.

**Plane Strain**

For the case of plane strain, the crack will be modelled as an *embedded crack* with the volume swept out by the unloading stress wave as a sphere of a radius r.

$$L = 2\eta\gamma a^2 - \frac{\sigma^2}{2E}\left(\frac{4}{3}\pi r^3\right) \quad (3.5)$$

Substituting for $r$ gives

$$L = 2\eta\gamma a^2 - \frac{\sigma^2 \pi a^3}{48E} \quad (3.6)$$

So that using the same substitution as before

$$\frac{da}{dn} = \frac{\Delta K^2}{48 K_{1c}^2} \quad (3.7)$$

where $K_{1c}^2$ is the critical strain intensity factor for plane strain.

Therefore, for both the cases of plane stress and plane strain fatigue of a perfectly elastic material with no crack tip plasticity, the fatigue life is dependent on $\Delta K^2$, but with differing constants. However, this condition is rarely found as most materials used under fatigue loading are ductile with crack tip plasticity.

**Elastic-Plastic Material: Plane Stress**

In an elastic-plastic material a plastic zone will form at the tip of the crack, an illustration of which can be seen in the Schardin photograph (Fig. 2). where the von Mises plastic lobes are evident. This plastic zone is constantly re-developed as the crack propagates in the time interval between the stress increasing to the critical stress and the return of the unloading stress wave. The magnitude of the unloading stress wave as it passes through the elastic-plastic boundary will be continuous. As the energy released from the body will increase as a function of the radius of the unloading radius cubed, and the stress decreases as an inverse function of the radius r, then the energy released will increase monotonically. It is assumed here that the energy released at any time can be expressed as a multiple (n) of the energy released by the unloading stress wave traversing the plastic zone.

The unloading stress wave will traverse both the elastic (K-dominated) stress field and plastic material (Fig. 2), and so the local stress $\sigma_*$ will be in the range $\sigma(fK) \leq \sigma_* \leq \sigma_y$, and hence in general it will be assumed that $\sigma_* = \zeta\sigma$.

Taking Irwin's estimate of the radius of the plastic zone under plane stress conditions as (Ewalds & Wanhill):

$$r_p = \frac{1}{2\pi}\left(\frac{\Delta K}{\sigma_y}\right)^2 \quad (3.8)$$

and taking the previously developed stress wave unloading model as:

$$V = \frac{\sigma_*^2}{2E}(vol) = \frac{\sigma_*^2}{2E}\left(nt\pi r_p^2\right) \quad (3.9)$$

the Lagrangian becomes:

$$L = 2\gamma\eta a^2 - \frac{n\zeta^2 t\sigma^6\left(\pi a^2\right)}{8\pi E \sigma_y^4} \quad (3.10)$$

Applying Euler-Lagrange gives:

$$\frac{\partial L}{\partial a} = 4\gamma\eta a - \left(\frac{\sigma}{\sigma_y}\right)^2 \frac{n\zeta^2 t \Delta K^2}{4E} = 0 \quad (3.11)$$

Letting $\eta a = \dfrac{da}{dn}$ gives:

$$\frac{da}{dn} = \frac{n\zeta^2 t}{8\pi}\left(\frac{\sigma}{\sigma_y}\right)^4 \frac{\Delta K^2}{K_c^2} \quad (3.12)$$

where $2\gamma E = K_c^2$

For a given fatigue test this can be written as:

$$\frac{da}{dn} = C\left(\frac{\sigma}{\sigma_y}\right)^4 \Delta K^2 \quad (3.13)$$

This is a major new result because the ratio of the maximum applied stress to the yield stress is identified for the first time as a highly significant variable. This ratio is called the Yield Stress Ratio (YSR) to distinguish it from the stress ratio $R$. This identification of the YSR is critical for answers to three of the problems with the Paris Law, as have been identified by many authors:
(a) the Paris law does not take into account the experimentally-observed dependence of fatigue crack growth rate on the stress ratio $R$ (this will be done later),
(b) the Paris Law does not account for the experimentally-observed linear dependence of the fatigue crack growth rate on the maximum stress intensity factor (again, done later), and
(c) why can the fatigue crack growth results be predicted based on a double stress intensity factor (both the maximum stress intensity factor and the normal stress intensity factor)?

What has happened over the years is that the questions outlined above have resulted in the constant in the Paris Law being manipulated to fit the experimental results without an overarching theoretical framework. This is now dealt with in detail below.

The previous equations are for the situation where the stress ratio $R = 0$ and so $\sigma = \Delta\sigma = \hat{\sigma}$. Hence equation (3.12) can equally well be written as:

$$\frac{da}{dn} = \frac{n\zeta^2 t}{8\pi}\left(\frac{\hat{\sigma}}{\sigma_y}\right)^4 \frac{\Delta K_{max}^2}{K_c^2} \qquad (3.14)$$

where $\Delta K_{max} = \hat{\sigma}\sqrt{\pi a}$ \qquad (3.15)

Hertzberg (Hertzberg 1976; Meneghetti 2012; Gavras, Lados et al. 2013) have shown that fatigue data are linearised if plotted as a function of the maximum value of the cyclic stress intensity factor $\Delta K_{max}$. This is now theoretically explained here for the first time through the influence of the newly identified YSR. There is a good argument to be advanced that it is the maximum stress intensity factor which is the most relevant parameter, as cracks tend to grow near the peak load in the cycle (Lindley, Palmer et al. 1978).

To consider now the effects of the stress ratio ($R$-ratio effects) $R = \dfrac{\check{\sigma}}{\hat{\sigma}}$:

$$\bar{\sigma} = \frac{\hat{\sigma} + \check{\sigma}}{2} = \frac{\hat{\sigma}}{2}\left(1 + \frac{\check{\sigma}}{\hat{\sigma}}\right) = \frac{\hat{\sigma}}{2}(1+R) = \hat{\sigma}(0.5 + 0.5R)$$

$$\hat{\sigma} = 2\bar{\sigma} = 2\hat{\sigma}(0.5 + 0.5R) \qquad (3.16)$$

$$\hat{\sigma} = 2\bar{\sigma} = \hat{\sigma}(1+R)$$

If an "effective stress intensity factor" is defined after the manner of Elber (Elber 1976) as:

$$\Delta K_{eff} = \Delta K(0.5 + 0.5R) \qquad (3.17)$$

then equation 3.14 can also be written as:

$$\frac{da}{dn} = \frac{n\zeta^2 t}{2\pi}\left(\frac{\hat{\sigma}}{\sigma_y}\right)^4 \frac{\Delta K_{eff}^2}{\sigma_y^2 K_c^2} \qquad (3.18)$$

This derivation is the first to show that a fatigue crack growth relationship can be derived from first principles including the stress-ratio ($R$-ratio effects). The $R$-ratio appears because of the manner in which the mean stress is interpreted, it is not of any fundamental significance in its own right. It also shows a strong dependence on the Yield Stress Ratio (YSR), which so far has not been detected for plane stress, but has been examined indirectly for plane strain (see later).

It should be noted that this is a theoretical derivation based on a simplified crack model – equation 3.18 has been shown not to apply in practice and requires a small modification, as shown later.

However, for a given test all the parameters in equation 3.18 are constant, except $\Delta K$, and so the dependence of the fatigue crack growth on the YSR will have been largely missed in fatigue testing standardised by the ASTM. They can be incorporated into the existing constant to give:

$$\frac{da}{dn} = C_1 \left(\frac{\hat{\sigma}}{\sigma_y}\right)^4 \Delta K_{eff}^2 \qquad (3.19)$$

where $C_1 = \left(\frac{n\zeta^2 t}{8\pi K_c^2}\right)$

Obviously the YSR will normally also be a constant for a given fatigue test, once the initial parameters have been set, but it is stated explicitly here to emphasise its presence.

**Plane Strain**

For the case of plane strain the crack will be modelled as an embedded crack with the volume swept out by the unloading stress wave as a sphere of a radius equivalent to a multiple (n) of the plastic zone radius. In this case Irwin's estimate of the plastic zone radius is (Ewalds and Wanhill):

$$r_p = \frac{1}{6\pi} \left(\frac{\Delta K}{\sigma_y}\right)^2 \qquad (3.20)$$

and so the Lagrangian becomes:

$$L = 2\gamma\eta a^2 - \frac{\sigma_*^2}{2E} \left(\frac{4}{3} n\pi r_p^3\right) \qquad (3.21)$$

Making all the substitutions as for plane stress above, and applying Euler-Lagrange gives:

$$\frac{\partial L}{\partial a} = 4\gamma\eta a - \frac{n\zeta^2 \sigma^8 \pi^2 a^2}{153\pi E \sigma_y^6} = 0 \qquad (3.22)$$

Letting $\eta a = \frac{da}{dn}$ and, in this case, $2\gamma E = K_{Ic}^2$, gives

$$\frac{da}{dn} = \frac{n\zeta^2}{306\pi}\left(\frac{\sigma}{\sigma_y}\right)^4 \frac{\Delta K^4}{\sigma_y^2 K_{1c}^2} \qquad (3.23)$$

which can be written as:

$$\frac{da}{dn} = C_2\left(\frac{\sigma}{\sigma_y}\right)^4 \Delta K^4 \qquad (3.24)$$

These equations are again for R = 0 and so $\sigma = \Delta\sigma = \hat{\sigma}$, hence equation (3.24) can equally well be written as:

$$\frac{da}{dn} = \frac{n\zeta^2}{306\pi}\left(\frac{\sigma}{\sigma_y}\right)^4 \frac{\Delta K_{max}^4}{\sigma_y^2 K_{1c}^2} \qquad (3.25)$$

which for a given test again linearises the results in terms of $\Delta K_{max}$, as originally found by Hertzberg (Hertzberg 1976).

When the stress-ratio substitutions are made equation 48 gives:

$$\frac{da}{dn} = \frac{n\zeta^2}{306\pi}\left(\frac{\hat{\sigma}}{\sigma_y}\right)^4 \frac{\Delta K_{eff}^4}{\sigma_y^2 K_{1c}^2} \qquad (3.26)$$

Alternatively, equation 3.26 can be written:

$$\frac{da}{dn} = C_2\left(\frac{\hat{\sigma}}{\sigma_y}\right)^4 \Delta K_{eff}^4 \qquad (3.27)$$

where $C_2 = \left(\dfrac{n\zeta^2}{306\pi\sigma_y^2 K_{1c}^2}\right)$

which is a constant for a given test.

If the YSR varies with the dimensionality of the stress wave unloading exponent, then in general this would suggest that:

$$\frac{da}{dn} = C\left(\frac{\hat{\sigma}}{\sigma_y}\right)^m \Delta K_{eff}^n \quad \text{or} \quad \frac{da}{dn} = C\left(\frac{\hat{\sigma}}{\sigma_y}\right)^m \Delta K^n \qquad (3.28)$$

so that the fatigue crack growth relationship can be written in terms of either $\Delta K_{max}^n$ or $\Delta K_{eff}^n$ or $\Delta K^n$, where it appears that the YSR may have a different exponent to the stress intensity factor.

As an aside, it was stated by Broek (Broek 1988) that no theoretical fatigue crack growth relationship can account for the stress ratio effect; this is incorrect. We may take a generic form of the Paris Law as:

$$\frac{da}{dn} = C\Delta K^n = C\left[(\hat{\sigma} - \breve{\sigma})\sqrt{\pi a}\right]^n = C\left[(1-R)\Delta K_{max}\right]^n \quad (3.29)$$

Therefore, the $R$-ratio effects have always been in the Paris Law, if it is rephrased as a function of the maximum stress intensity factor.

A further effect that can be explained by this derivation is the experimental correlation of fatigue crack growth rate with the combined ($\Delta K.\Delta K_{max}$) formulation, which has been found experimentally (Meneghetti 2012; Gavras, Lados et al. 2013). Thus, from equation 3.28:

$$\frac{da}{dn} = C_2\left(\frac{\sigma}{\sigma_y}\right)^4 \Delta K^n = C_2\left(\frac{\sigma}{\sigma_y}\right)^2 \Delta K^{\frac{n}{2}}\Delta K_{max}^{\frac{n}{2}} = C_2\left(\frac{\sigma}{\sigma_y}\right)^2 \Delta K^\alpha \Delta K_{max}^{n-\alpha} \quad (3.30)$$

**Part IV  Discussion**

Elber (Elber 1976) found from experimental trials that:

$$\Delta K_{eff} = \Delta K(0.5 + 0.4R) \quad (4.1)$$

was a good experimental fit to his data for 2024 aluminium, whereas Schijve and Broek (Schijve and Broek 1962) found that a second order model was a slightly better fit for 2024-T3. Both of these experimental estimates, within the scatter of the results, are in good agreement with equation (4.1) and so it is concluded that this derivation of a theoretical fatigue crack growth relationship includes mean stress effects which agree with the experimental data for the first time.

It is frequently said in the literature that the use of the $R$-ratio uniquely defines a fatigue loading cycle; this is incorrect. However, as $R = \frac{\breve{\sigma}}{\hat{\sigma}}$ has two degrees of freedom the stresses themselves are indeterminate. That is, to correctly and completely specify a fatigue loading cycle it is necessary to give both $\breve{\sigma}$ $and$ $\hat{\sigma}$ as fractions of $\sigma_y$, and in absolute terms. It is considered that this mistake has led to much confusion over the years.

Brown (Brown 1988) has shown that, even if the stress ratio $R$ is kept constant, as the applied stress is increased the fatigue crack growth relationships diverge (Fig. 4.1), indicating that other factors are operating.

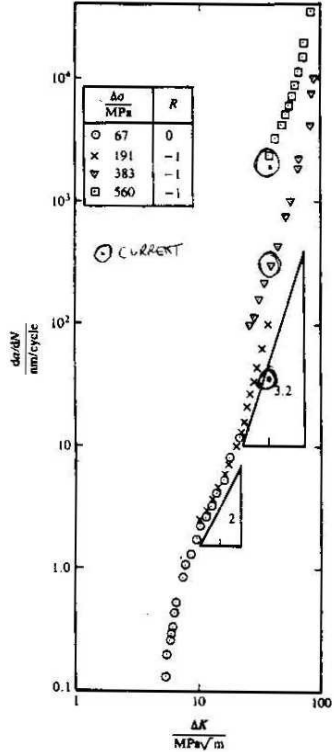

Fig. 4.1 Correlation with Brown's results (Brown 1988)

Although Brown's curves hardly overlap, if equation 3.23 is normalised at:

$$\Delta\sigma = 191 MPa$$
$$\Delta K = 30 MPa\sqrt{m}$$
$$\sigma_y = 395 Mpa$$

and then applied to the other loading conditions.  It can be seen that accounting for the yield stress ratio according to equation 3.23 accounts for the divergence of the experimental results to a reasonable degree of accuracy, and the results can be best described by:

$$\frac{da}{dn} = C_2 \left(\frac{\hat{\sigma}}{\sigma_y}\right)^{3.98} \Delta K^{3.2} \qquad (4.2)$$

as for $R = -1$  $\hat{\sigma} = \frac{\Delta\sigma}{2}$

Nisitani and Goto (Nisitani and Goto 1986) explored fatigue crack growth rates over wide ranges of applied stress. They did not use a stress intensity factor approach, but simply plotted their results as a function of applied stress and crack length.  They proposed a "Small Crack Growth Law" as:

$$\frac{da}{dn} = C\sigma^\xi a \qquad (4.3)$$

and found that $4 < \xi < 8$ for a variety of steels. From the models proposed here, expanding equation 3.13 for plane stress gives:

$$\frac{da}{dn} \propto \sigma^4 a \tag{4.4}$$

whereas equation 3.24 for plane strain gives:

$$\frac{da}{dn} \propto \sigma^8 a^2 \tag{4.5}$$

Due to the small range of crack lengths considered by Nisitani and Goto, the resolution on crack length is very small, and their results for σ versus crack length 'a' show a concave upwards relationship, that could equally well fit an $a^2$ relationship.

Analysing their data for 0.45% carbon steel with a lower yield value of 364 MPa, in a necked specimen, which is taken as an embedded crack, i.e. a plane strain model, and normalising on their mid-range results at $\Delta\sigma = 480 MPa$, $a = 1.0 mm$, equation 4.5 becomes:

$$\frac{da}{dn} = (3.6e - 6)\sigma^8 a^2 \tag{4.6}$$

which when fitted to Nisitani and Goto's results over the full range of the experimental variables, as shown in Fig. 4.2, is seen to be a good fit.

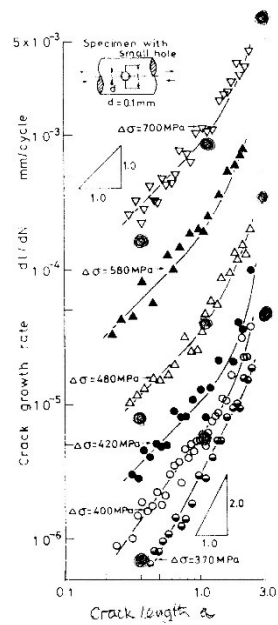

Fig. 4.2 Nisitani & Goto's results (Nisitani and Goto 1986)

Amsterdam and Grooteman (Amsterdam and Grooteman 2016) explored the fatigue of aluminium 7075-T7351 using a test where the crack length varied from 9.8 to 40.8 mm, and changed from plane strain to plane stress conditions as it propagated. They found that the Paris exponent was lower for plane stress (around 2.5) than for plane strain (around 3.6), which is in broad agreement with the model proposed here, being influenced by the size and shape of the plastic zone (stress state) and the dimensionality of the unloading stress wave.

The form of equations 3.13 and 3.24 also throw light on the phenomenon of the "fatigue threshold", which is determined by "load shedding" where the length of the fatigue crack is essentially kept constant, the load reduced and the fatigue crack propagation rate determined. This is in marked contrast to normal fatigue testing where the load is kept constant and $\Delta K$ varied by the crack growing. Therefore, there are in effect two different types of fatigue test: one where the load (hence stress) is kept constant and the crack length varies, and one where the crack length is constant and the load varies. To explore this further take equation 3.24, where the slope can be calculated for both cases outlined above, giving for constant applied stress and a growing crack:

$$\frac{\partial \left(\frac{da}{dn}\right)}{\partial a} \propto \sigma^6 a \propto \Delta K^2 \tag{4.7}$$

and for load shedding:

$$\frac{\partial \left(\frac{da}{dn}\right)}{\partial \sigma} \propto \left(\frac{\sigma}{\sigma_y}\right)^7 a^2 \propto \left(\frac{\sigma}{\sigma_y}\right)^3 \Delta K^4 \tag{4.8}$$

It can be seen clearly that the slope for equation 4.8 is significantly larger than that indicated by equation 4.8. It is considered that the principal effect during load shedding is due to the yield stress ratio. Furthermore, this data does not show a "sigmoidal trend" as is often claimed. They are due to three entirely different types of fatigue test (load shedding, crack growth elongation, and monotonic fracture) plotted on the same axes, as shown schematically in Fig. 4.3.

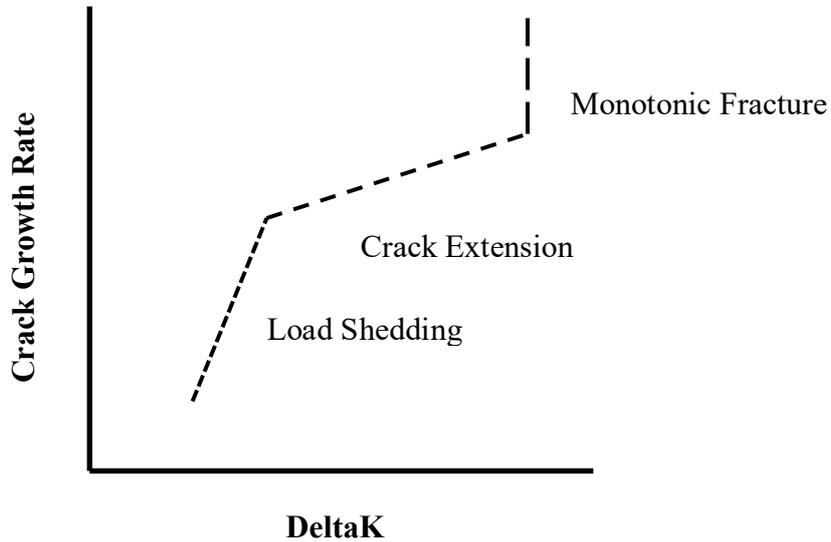

Fig. 4.3 Schematic of the three types of fatigue test

Seetharam and Dash (Seetharam and Dash 1991) are one of the very few investigators (indeed, the only one found!) who have produced a set of fatigue crack growth curves where the full range of variables were noted. Their results are shown below.

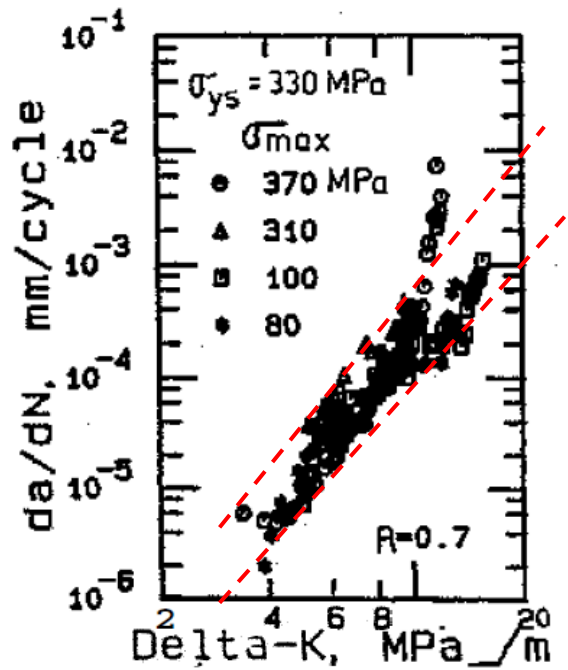

Fig. 4.3 Seetharam & Dash's results (Seetharam and Dash 1991)

These results clearly demonstrate that they are strongly influenced by the YSR. However, looking at the superimposed straight line relationships, they are best described as:

$$\frac{da}{dn} = 3.25e^{-8}\left(\frac{\hat{\sigma}}{\sigma_y}\right)\Delta K^4 \tag{4.9}$$

where the exponent on the YSR is only 1, and not 4 as predicted by the theory. Therefore, in general it may be that the generic form is:

$$\frac{da}{dn} = C\left(\frac{\hat{\sigma}}{\sigma_y}\right)^m \Delta K^n \tag{4.10}$$

What is very clear from this work is that the effect of the maximum stress in the fatigue cycle, and what fraction it is of the yield stress, has been largely overlooked in fatigue research.

**Conclusion**

The above results may be summarised as follows:
1) The monotonic Griffith crack has been re-analysed using a Lagrangian, and the solutions found to agree with Griffith's analysis.
2) A stress wave unloading model has been introduced, and shown to give the Griffith-Irwin solution without the need to invoke two separate equations.
3) A Lagrangian has also been applied to the process of fast fracture using Mott's inertial correction. A formulation of the dynamic stress intensity factor, as a function of the crack velocity has been derived, and is in good agreement with experiments.
4) A Lagrangian has also been applied to fatigue of perfectly elastic material under both plane stress and plane strain conditions. In both cases the fatigue relationship is proportional to $\Delta K^2$.
5) When crack tip plasticity is incorporated, the equations show a marked dependence on the ratio of the applied stress to the yield stress, called here the Yield Stress Ratio (YSR).
6) The form of the fatigue crack growth relationship derived here including the Yield Stress Ratio effects, can also be expressed as a function of the '$R$-ratio', and the "effective stress intensity factor" ($\Delta K_{eff}$). The results are again in accord with experimental evidence.
7) The results show that the fatigue crack growth rates can also be linearised with respect to $\Delta K_{max}$, and a combination of ($\Delta K_{max}$ and $\Delta K$) as a function of the maximum stress in the cycle, as has been found experimentally.
8) It is considered that the Yield Stress Ratio effect is the predominant reason why an unmodified Paris Law cannot account for the fatigue crack growth rate over large ranges of the applied stress. Initial experimental confirmations of the validity of the equations derived are given.
9) The fatigue crack growth rate including the Yield Stress Ratio (YSR) can also account for the significant change of slope observed in "load shedding" tests, as used to find the fatigue threshold.
10) The fatigue crack growth rate plotted against large ranges of $\Delta K$ using both load shedding and constant load tests through to monotonic fracture are not a sigmoidal curve, but three different crack growth regimes plotted on one set of axes.

The overall conclusion of this work is that, lacking a formal structure and the identification of the YSR as a dominant parameter, many arbitrary manipulations of the constant in the Paris Law have taken place, resulting in much confusion.